\documentclass[useAMS,usenatbib]{mn2e}
\input psfig.sty
\usepackage[dvips]{color}
\usepackage{graphicx}
\usepackage{epsfig}

\title[Large scale structure and ISW effect in decaying vacuum cosmology]
  {Large scale structure and ISW effect in decaying vacuum cosmology}
\author[H. Velten, H. A. Borges, S. Carneiro, R. Fazolo and S. Gomes]{H. Velten$^{1,2}$\thanks{E-mail:
velten@pq.cnpq.br}, H. A. Borges$^{3,4}$, S. Carneiro$^3$, R. Fazolo$^1$ and S. Gomes$^1$ \\
$^1$Departamento de F\'isica, Universidade Federal do Esp\'irito Santo, Vit\'oria, ES, Brazil\\$^2$CPT, Aix Marseille Universit\'e, UMR 7332, 13288 Marseille, France\\$^3$Instituto de F\'{\i}sica, Universidade Federal da Bahia, 40210-340, Salvador, BA, Brazil\\$^4$Institute of Cosmology and Gravitation, University of Portsmouth, PO1 3FX, Portsmouth, UK}
\begin{document}
\date{\today}

\pagerange{\pageref{firstpage}--\pageref{lastpage}} 

\def\LaTeX{L\kern-.36em\raise.3ex\hbox{a}\kern-.15em
    T\kern-.1667em\lower.7ex\hbox{E}\kern-.125emX}
\label{firstpage}

\maketitle

\begin{abstract}
The concordance particle creation model - a class of $\Lambda(t)$CDM cosmologies - is studied using large scale structure (LSS) formation, with particular attention to the integrated Sachs-Wolfe (ISW) effect. The evolution of the gravitational potential and the amplitude of the cross-correlation of the cosmic microwave background (CMB) signal with LSS surveys are calculated in detail. We properly include in our analysis the peculiarities involving the baryonic dynamics of the $\Lambda(t)$CDM model which were not included in previous works. Although both the $\Lambda(t)$CDM and the standard cosmology are in agreement with available data for the CMB-LSS correlation, the former presents a slightly higher signal which can be identified with future data.\end{abstract}


\section{Introduction}
\label{intro}

Although the great success of the standard $\Lambda$CDM cosmological model in describing most observations, we are still distant from the full understanding of the cosmic dynamics. Apart from the so called small scale problems in the dark matter sector, there is also a huge discussion about dark energy properties. At the same time, the efficient approach given by a cosmological constant $\Lambda$ still seems to face some challenges, mainly from the theoretical point of view.

Among the viable alternatives, it has been argued that a model with the mechanism of dark matter particle production at a constant rate $\Gamma$ (implying in a dynamical vacuum term $\Lambda(t)$) is also capable to produce a concordance model \citep{LtCDM2012}, although with a larger current matter fraction $\Omega_{m0} \approx 0.45$\footnote{For any component $i$ we define $\Omega_i=\rho_i/\rho_{c0}$ with $\rho_{c0}=3 H^2_0 / 8\pi G$, where the subscript $0$ denotes today's values. In this paper we will take $8\pi G = c = 1$.}.

If the total energy is given by $\rho=\rho_m +\Lambda$ and the pressure of the vacuum contribution sets to $p_{\Lambda}=-\Lambda$, a constant $\Gamma$ is equivalent to take $\Lambda=2\Gamma H$, where $H=\dot{a}/a$ is the Hubble rate\footnote{Strictly speaking, this equivalence is only valid if we neglect the conserved baryons in the background equations. In the presence of a small baryonic content, taking $\Lambda \propto H$ does not lead to an exactly constant $\Gamma$.}. Taking today's values one finds $\Gamma = 3 \Omega_{\Lambda 0} H_0/2$. The matter dynamics corresponds to the balance
\begin{equation}\label{contrhom}
\dot{\rho}_{m}+3H\rho_m=-\dot{\Lambda},
\end{equation}
from which it is not difficult to show that $\dot{\Lambda}=-\Gamma \rho_{m}$. The background expansion is given by
\begin{equation}\label{Ht}
H=H_0 \left( 1-\Omega_{m0}+\Omega_{m0}a^{-3/2}\right),
\end{equation}
from which we can derive the matter density evolution
\begin{equation} \label{Omega}
\rho_m =3H_0^2 \left[ \Omega_{m0}^2 a^{-3}+(1 - \Omega_{m0}) \Omega_{m0}a^{-3/2}\right].
\end{equation}
Note that, as in the flat $\Lambda$CDM model, there are here also only two free parameters.
The concordance of this $\Lambda(t)$ cosmology with $\Omega_{m0} \approx 0.45$ has been verified via many different observational data at both the background \citep{LtCDMb2006,LtCDMb2006b,LtCDMb2006c,Velten} and perturbative \citep{LtCDMbp2008,LtCDMp2011,LtCDMp2015} levels. 

Concerning structure formation, the full CMB spectrum has not yet been obtained for this model, but CMB physics can also be accessed via ISW studies, i.e., how time-varying gravitational potential wells change the temperature of the CMB photons as they cross structures \citep{ISW}. The expansion of an Einstein-de Sitter universe compensates the clustering of structures, producing no ISW effect, i.e., in a matter dominated universe there is no ``late time'' ISW effect\footnote{The ``early time'' ISW effect is related to non-insignificant radiation density just after photon decoupling.}. Dark energy modifies the background expansion and leads to a net contribution to $(\Delta T/T)^{ISW}$. Then, it is expected that the modified background and perturbative expansion of the $\Lambda(t)$CDM model leaves a distinct imprint on the ISW signal.

Since the ISW is a secondary CMB temperature effect its detection occurs only via the cross-correlation with other large scale probes like galaxies and quasars surveys \citep{Cross}. Ref. \citep{wang} used the cross-correlation technique to probe the $\Lambda(t)$CDM model, finding an increase of the ISW-galaxy spectrum ($C^{Tg}_l$) in comparison to the $\Lambda$CDM model. Our aim in this paper is to perform this analysis taking care with some peculiarities of the model. In particular, it is important to
differentiate the evolution of the baryonic and dark matter components. At the background level, 
baryons are included in the conserved part of (\ref{Omega}). Concerning perturbations, the observed matter power spectrum $P(k)=\left|\delta(k)\right|^2$ is a measurement of the baryonic clustering, which is sourced by the gravitational potential. In the $\Lambda$CDM model there is almost no difference between the evolution on large scales of such components. The same does not happen in the $\Lambda(t)$CDM model, where one observes a late-time suppression in the dark matter and total matter contrasts owing to dark matter production, a suppression not observed in the baryonic contrast.

Our interest is also justified by a possible tension between the theoretical $\Lambda$CDM predictions and the observed ISW signal. Some analysis have reported a cross-correlation signal $1\sigma$-$2\sigma$ above that expected in the standard cosmology \citep{crossincrease,crossincreaseb} (see \citep{giann} for a critical review on this issue). Recent studies claim a better statistical concordance though the observed signal is still higher than the theoretical one \citep{crossok, Spergel}. Also, the inferred stacking of CMB data on the position of superstructures is about 5 times larger than the $\Lambda$CDM prediction \citep{stack,stackb,stackc,stackd}. This result has recently been confirmed by the 2015 release of the Planck CMB data \citep{PlanckISW}.

In the next section we develop the background expansion which will be used in this paper. In section \ref{sectionIII} we explore the scalar perturbations of the model. The connection with the ISW effect is made via a direct calculation of the evolution of the gravitational potential (\ref{pot}) and the ISW-LSS cross-correlation (\ref{crossISWLSS}). We conclude in the final section.

\section{$\Lambda(t)$ background dynamics}

The brief description of the $\Lambda(t)$CDM dynamics given in equations \ref{contrhom}-\ref{Omega} has been widely derived in the literature (see \citep{LtCDM2012} and references therein). Note that as in the standard case the important quantity is the total matter density $\Omega_{m0}$, the sum of the dark matter $\Omega_{dm0}$ and baryons $\Omega_{b0}$. The latter is constrained by nucleosynthesis and is well approximated by $\Omega_{b0}=0.05$. We assume hereafter this value. The baryonic sector is conserved and therefore we have the continuity equations 
\begin{equation}
\dot{\rho}_{dm}+3H\rho_{dm}=-\dot{\Lambda}
\end{equation}
and 
\begin{equation}
\dot{\rho}_{b}+3H\rho_{b}=0.
\end{equation}
Adding these equations we obtain the total matter conservation equation (\ref{contrhom}). With $\Lambda=2\Gamma H$, the dynamics presented in section \ref{intro} follows. That is, the background expansion is obtained after solving the system given by the continuity equation 
\begin{equation}
aH\rho^{\prime}_{m}+3H \rho_{m}=\Gamma \rho_{m}
\end{equation}
and the Friedmann equation
\begin{equation}
2aHH^{\prime} + \rho_{m}=0,
\end{equation}
where the prime means derivative with respect to the scale factor. 

\section{Perturbative dynamics}\label{sectionIII}

\subsection{Growth functions}

When cosmological perturbation theory is used to study the large scale structure of the universe we are most interested in the evolution of scalar quantities like the density contrast and the gravitational potential. The $\Lambda(t)$CDM models have the particularity of suppressing the growth of dark matter contrast $\delta_{dm}$. However, the observed structures reflect the behavior of the baryonic contrast, $\delta_b=\delta\rho_b / \rho_b$, instead. In the standard cosmology this quantity coincides with the dark matter contrast, and the total matter contrast $\delta_m$ denotes both behaviors. For the $\Lambda(t)$CDM model, however, we should make a distinction between the different components. Using the fact that the vacuum fluctuations are negligible \citep{LtCDMp2011} and choosing the comoving synchronous gauge we obtain the set of equations\footnote{A different approach to the baryonic sector can be found in \citep{Zimdahl}.}
\begin{equation}
\dot{\delta}_{m}+{\Gamma}\delta_{m}=\frac{\dot{h}}{2},
\end{equation}
\begin{equation}
\dot{\delta}_{b}=\frac{\dot{h}}{2},
\end{equation}
\begin{equation}
\ddot{h}+2H\dot{h}=\rho_{m} \delta_{m},
\end{equation}
where $h$ is the trace of the spatial metric perturbation. Combining these equations and changing to the derivative with respect to the scale factor we find
\begin{eqnarray}\label{eqdeltadm}
a^2H^2 \delta^{\prime\prime}_{m}+aH(3H+aH^{\prime}+{\Gamma})\delta^{\prime}_{m}+
\nonumber \\
2{\Gamma} H \delta_{m} = \frac{1}{2} \rho_{m} \delta_{m}
\end{eqnarray}
and
\begin{eqnarray}\label{eqdeltab}
a^2H^2 \delta^{\prime\prime}_{b}+aH(3H+aH^{\prime})\delta^{\prime}_{b} =\frac{1}{2}\rho_{m} \delta_{m}.
\end{eqnarray}

\begin{figure}
\begin{center}
\includegraphics[width=0.35\textwidth]{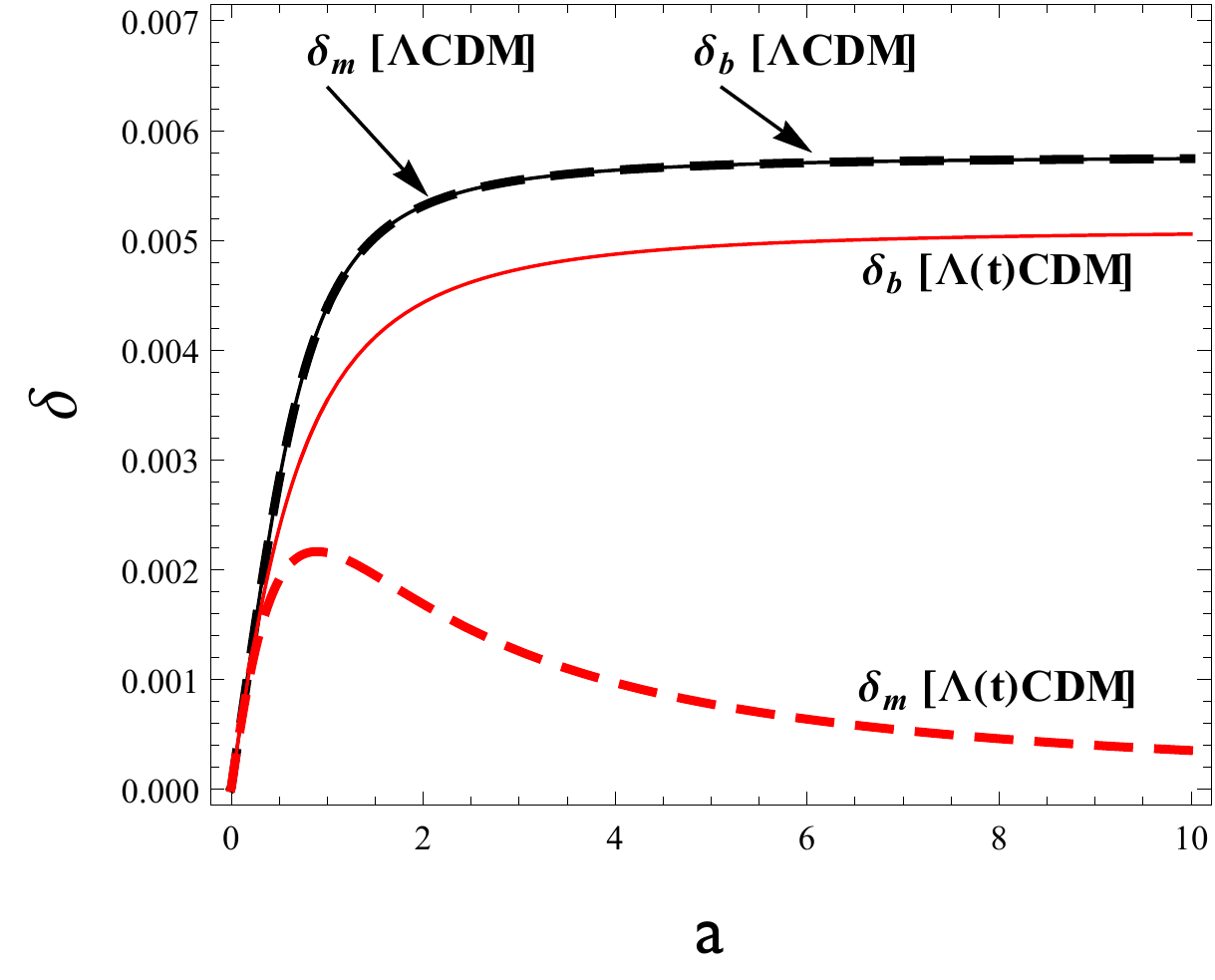}
\caption{Evolution of the linear density contrasts. The black curves correspond to the baryonic (solid-black curve) growth and total matter (dashed-black curve) growth for the $\Lambda$CDM model with $\Omega_{m0} = 0.23$. Related plots for the $\Lambda(t)$CDM with $\Omega_{m0}=0.45$ are in solid red (baryons) and dashed red (total matter), respectively.}
\label{deltas}
\end{center}
\end{figure}

In Fig. \ref{deltas} we plot the evolution of the density contrasts as calculated in (\ref{eqdeltadm}) and (\ref{eqdeltab}). For the $\Lambda$CDM model (black curves) we set $\Omega_{m0}=0.23$, the best-fit of LSS observations \citep{2dF,SDSS}\footnote{Using, instead, the $\Lambda$CDM concordance value $\Omega_{m0}=0.27$ does not change significantly our final results.}. We show the $\Lambda (t)$CDM evolutions for $\delta_b$ and $\delta_m$ in the solid red and dashed red curves, respectively. We have used initial conditions $\delta_{b}(a=0.001)=\delta_{m}(a=0.001)=10^{-5}$ and $\delta^{\prime}_{b}(a=0.001)=\delta^{\prime}_{m}(a=0.001)=0$. As expected, $\delta_b$ and $\delta_{m}$ have the same evolution in the standard case. On the other hand, the plot shows a late-time growth suppression of $\delta_{m}$ in the $\Lambda (t)$CDM model, a consequence of dark matter creation. The same suppression is not observed in the baryonic contrast though its amplitude achieves a plateau on late times which is slightly below the standard scenario.

\subsection{Evolution of the gravitational potential}\label{pot}

The evolution of the gravitational potential $\Phi$ determines the integrated Sachs-Wolfe contribution to the total CMB spectrum via the formula
\begin{equation}\label{ISWdeltaT}
\frac{\Delta T}{T}^{ISW} = 2\int^{1}_{a_d} \frac{\partial \Phi}{\partial a} da,
\end{equation}
where the subscript ``d'' denotes the decoupling time, and $a =1$ at present. Writing down the Poisson equation in the comoving synchronous gauge we find
\begin{equation} \label{EqPoisson}
k^2 \Phi= - \frac{a^2}{2}\rho_{m}\delta_{m},
\end{equation}
where $k$ is the comoving wave-number. This expression allows us to calculate the ISW, which can be computed via the line of sight integration (\ref{ISWdeltaT}). It is useful to perform it in terms of the comoving distance
\begin{equation}
\chi(a)=\int^1_a \frac{da}{a^2 H(a)}.
\end{equation}
The ISW then reads
\begin{eqnarray}\label{deltaTT}
\frac{\Delta T}{T}^{ISW}=\frac{\delta_{m}(a=1)}{k^2}\int^{\chi_d}_0 a^2 H(a) \frac{dQ_{m} (a)}{da} d\chi,
\end{eqnarray}
where we have defined
\begin{eqnarray}\label{defQ}
Q_{m}(a) &=& a^2 \rho_{m}(a) D^+_{m}(a),
\end{eqnarray}
with
\begin{eqnarray}\label{Dplusg}
D^+_{m}(a) &=& \frac{\delta_m(a)}{\delta_{m}(a=1)}.
\end{eqnarray}

\begin{figure}
\begin{center}
\includegraphics[width=0.35\textwidth]{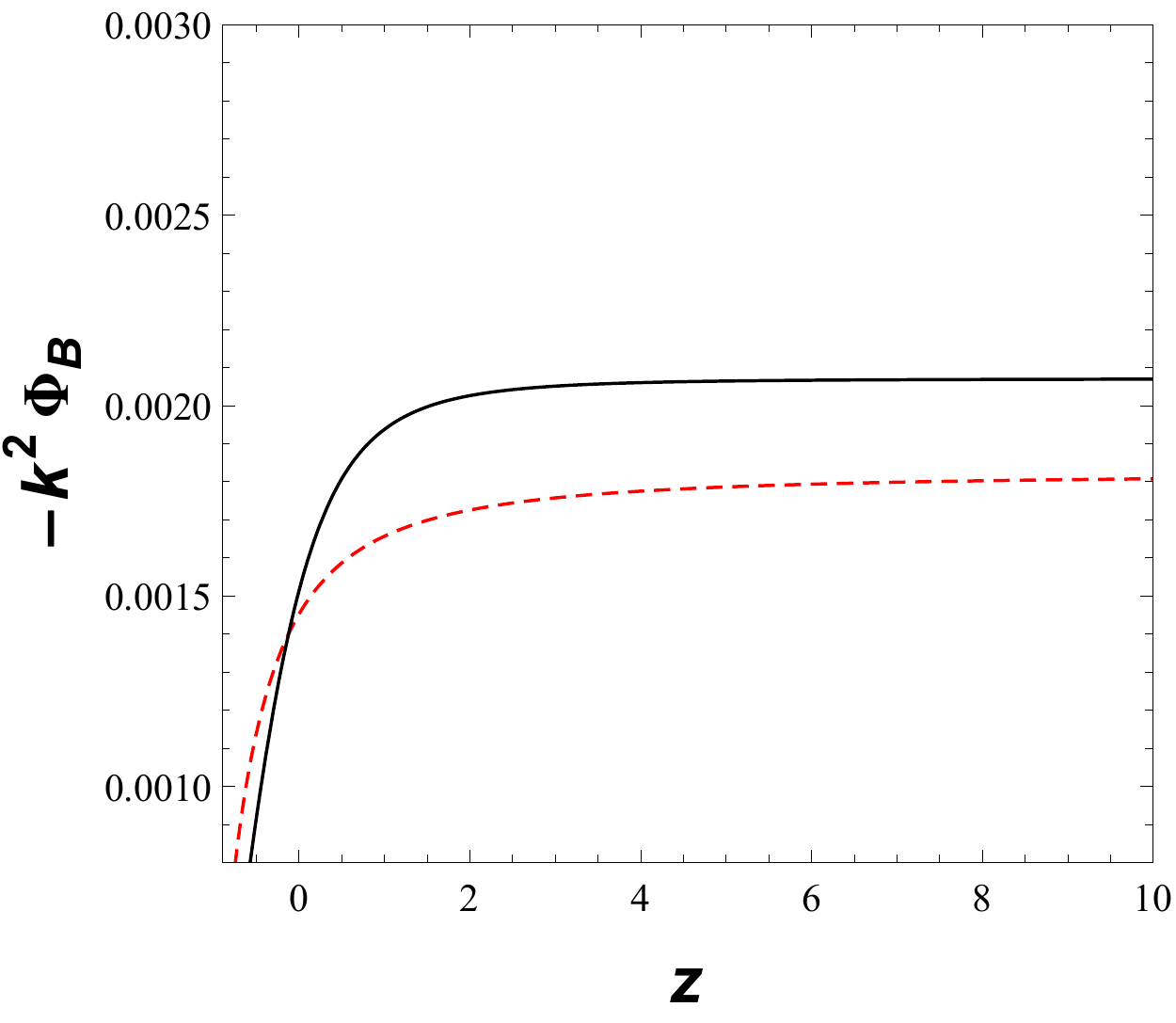}
\caption{Evolution of the gravitational potential for the $\Lambda$CDM model with $\Omega_{m0} = 0.23$ (black) and for the $\Lambda(t)$CDM with $\Omega_{m0} = 0.45$ (red-dashed).}
\label{Figpotential}
\end{center}
\end{figure}

The baryonic and total matter density contrasts can be directly calculated from the solution of Eqs. (\ref{eqdeltadm}) and (\ref{eqdeltab}). Using Eq. (\ref{EqPoisson}), we compare in Fig. \ref{Figpotential} the predictions for the $\Lambda$CDM and $\Lambda (t)$CDM gravitational potentials as functions of the redshift $z$. We access the observational predictions for the ISW effect via the cross-correlation of CMB maps and LSS surveys, to be done in next section.

\subsection{Correlating CMB maps and galaxy surveys}\label{crossISWLSS}

In order to compute the cross-correlation  between CMB and LSS we also need to describe the evolution of the observed galaxy contrast $\delta_g$ on the line of sight. This quantity depends on the survey design and is obtained from
\begin{equation}\label{deltagal}
\delta_g = \int^{\chi_d}_0 b(z) \frac{dN}{dz} H(a) \delta_{b}(a) d\chi,
\end{equation}
where $\delta_{b}$ is the baryonic linear contrast, and $b(z)$ is the standard bias between $\delta_g$ and $\delta_b$\footnote{In the $\Lambda$CDM model is irrespective to use $\delta_b$ or $\delta_m$ in (\ref{deltagal}), but the same is not true for the $\Lambda(t)$ case, as can be seen in Fig. \ref{deltas}.}.

The redshift distribution of the observed galaxy sample is a model-independent quantity. Each survey has its own $dN/dz$ histogram function. We use in this work data from the NRAO VLA Sky Survey (NVSS)\footnote{http://www.cv.nrao.edu/nvss/} and the Wide-field Infrared Survey (WISE)\footnote{http://wise.ssl.berkeley.edu/} catalogues. Both surveys are widely used for cross-correlation studies. The NVSS covers the entire north sky of $-40$ deg declination in one band. The produced catalogue of discrete sources contains more than one million objects. Full details appear in Ref. \citep{Condon}. WISE has an entire sky scanning strategy in four
frequency bands. It has an average redshift $0.3$, but reaching up to $z \sim 1$, collecting about 500 million sources which are in general not restricted to point-like objects such as, for example, stars and unresolved galaxies. The data used in this work is taken from the analysis recently performed in Ref. (\citep{Spergel}) which uses a larger sample by applying less-conservative cuts to the dataset in comparison to previous works \citep{
crossincreaseb, crossok}.

For the NVSS catalogue one has 
\begin{equation}
b(z)\frac{dN}{dz}=b_{eff}\frac{\alpha^{\alpha+1}}{z_{\star}^{\alpha+1}\Gamma(\alpha)}z^{\alpha} {\rm exp} \left[-\frac{\alpha z}{z_{\star}}\right],
\end{equation}
where the values $b_{eff}=1.98$, $z_{\star}=0.79$ and $\alpha=1.18$ were fitted in \citep{Ho}.
For the WISE catalogue $dN/dz$ can be obtained numerically from Ref. \citep{wise} and, following Ref. \citep{Spergel}, we adopt a constant bias $b(z)=1.41$. We will limit our results to constant bias models since the use of time-dependent $b(z)$ bias will not change our main conclusion.

Combining (\ref{deltaTT}) and (\ref{deltagal}), the multipole coefficients for the cross-correlation spectrum are given by
\begin{equation}
C_{Tg}(l) =  \int^{a_d}_{1} \frac{W_T (a)\, W_g (a)}{a^2 H(a)} \frac{P(k=l/\chi)}{l^2} da,
\end{equation}
where we have defined the weight functions
\begin{eqnarray}
W_T\,(a) &=& a^2 H(a) \frac{dQ(a)}{da}, \\
W_g\,(a) &=& H(a) b(z)\frac{dN}{dz} D^+_{b}(a),
\end{eqnarray}
with
\begin{equation}
D^+_{b}(a) = \frac{\delta_b(a)}{\delta_{b}(a=1)}.
\end{equation}
We have also defined the crossed power spectrum
\begin{equation}
P(k) = \frac{\delta_m(a=1)}{\delta_b(a=1)} P_b(k),
\end{equation}
where $P_b(k)$ is the observed baryonic spectrum, given by
\begin{equation}
P_b(k) = P_0 k^{n_s} T^2(k),
\end{equation}
where $P_0$ is a normalisation constant determined from observations \citep{LtCDMbp2008}. This normalisation constant is not the same as in the $\Lambda$CDM model as can be seen by an inspection of figures \ref{deltas} and \ref{Figpotential}. It can also be determined by writing $P_0 \delta_b^2(1) = \bar{P}_0 \bar{\delta}_b^2(1)$, where the barred quantities are the standard ones \citep{LtCDMp2015}. For $T(k)$ we will use the BBKS transfer function \citep{Bardeen1986,Martin} which, for $\Omega_{b0}\ll\Omega_{m0}$, can be approximated by
\begin{eqnarray}
&&T(x=k/k_{eq})=\frac{ln[1+0.171x]}{(0.171x)}\times \\ \nonumber
&&\left[1 + 0.284x + (1.18x)^2 + (0.399x)^3 + (0.490x)^4\right]^{-0.25}.
\end{eqnarray}
Here, $k_{eq}^{-1}$ is the comoving Hubble radius at the time of matter-radiation equality. From (\ref{Omega}) it is easy to show that it is given by \citep{LtCDMbp2008,LtCDMp2015} $k_{eq} = 0.073 h^2 \Omega_{m0}^2$ Mpc$^{-1}$, where, here, $h = H_0/(100$ km/s-Mpc) and we have set the present radiation density as $\Omega_{R0} = 4.15 \times 10^{-5} h^{-2}$. We will adopt $h=0.7$ and $n_s = 1$.

\begin{figure}
\begin{center}
\includegraphics[width=0.35\textwidth]{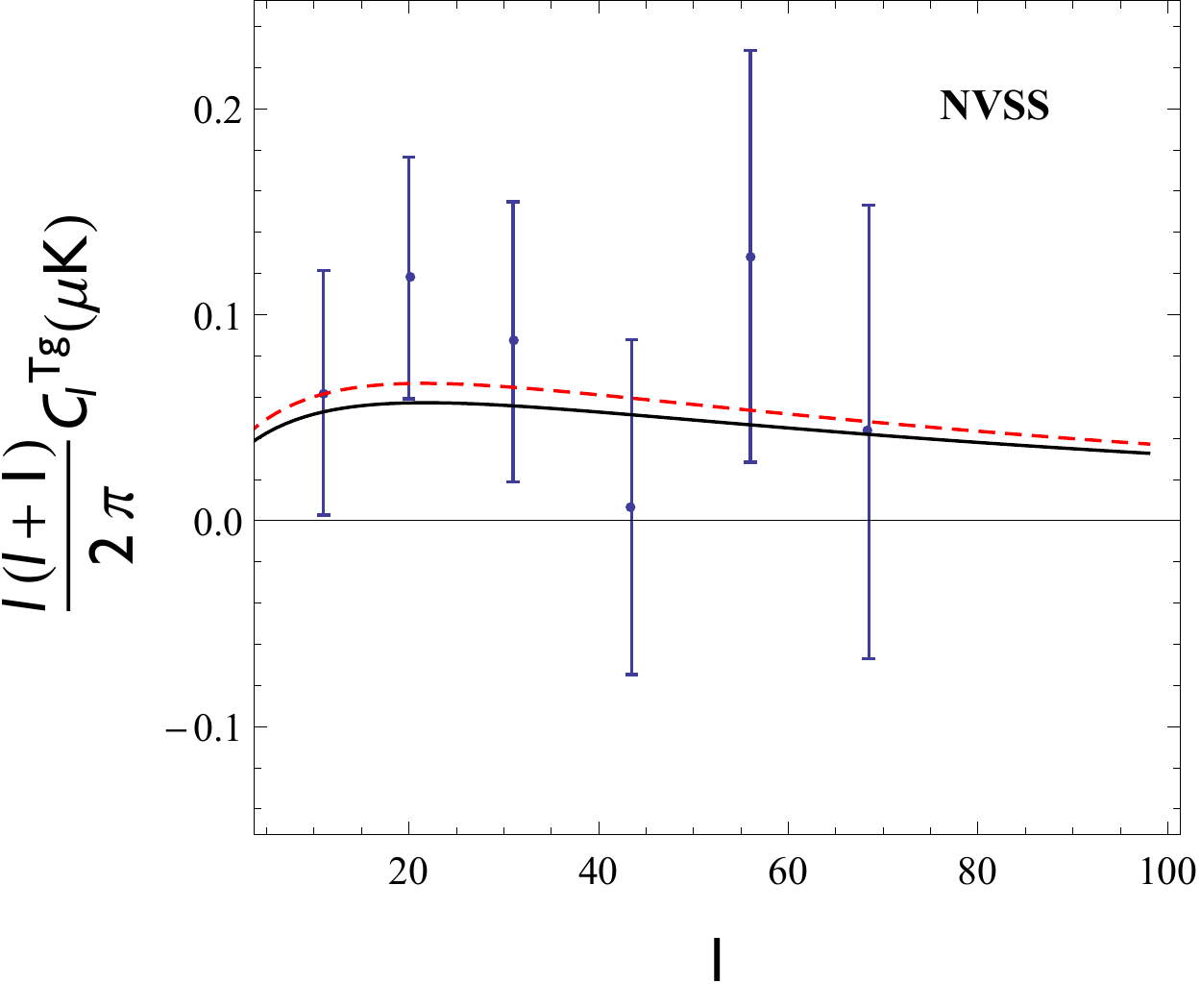}
\caption{CMB-LSS cross-correlation spectrum using the NVSS data taken from \citep{Ho} (see text for details). The black line represents the $\Lambda$CDM with $\Omega_{m0}=0.23$. The red-dashed line is our inferred spectrum for the $\Lambda$(t)CDM model using $\Omega_{m0}=0.45$.}
\label{fig3}
\end{center}
\end{figure}

\begin{figure}
\begin{center}
\includegraphics[width=0.35\textwidth]{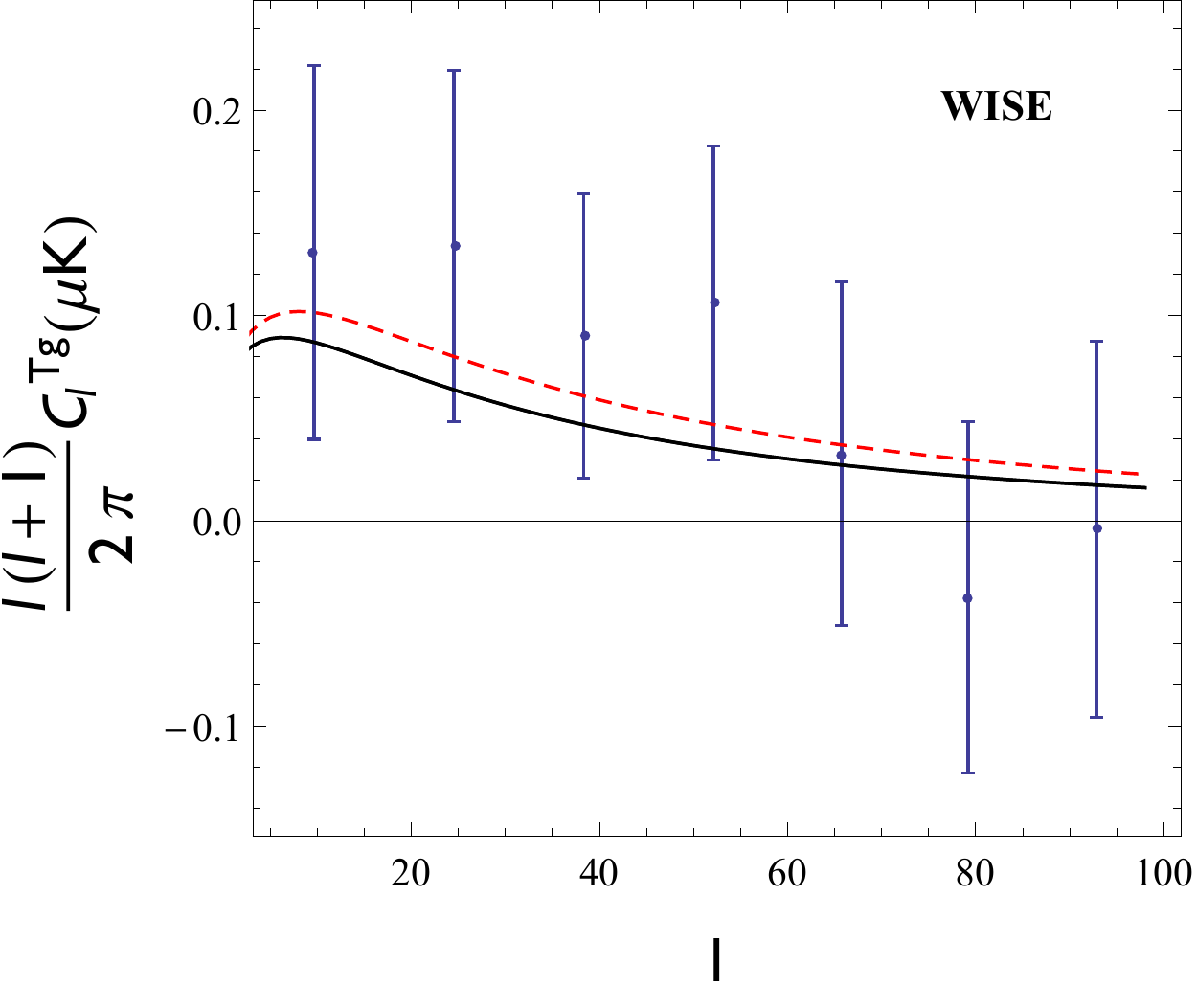}
\caption{The same as Fig. \ref{fig3} for the WISE data taken from \citep{Spergel}.}
\label{fig3b}
\end{center}
\end{figure}

The resulting cross-correlation spectrum is show in Figs. \ref{fig3} and \ref{fig3b}. For the NVSS data (Fig. \ref{fig3}) we follow the presentation of data used in \citep{wang}, where the original data points of Ref. \citep{Ho} located at $l<10$ and $l>70$ have been removed due to their high dispersion. The WISE data presented in Fig. \ref{fig3b} is taken from \citep{Spergel}. In both cases, the $C^{Tg}_l$ spectrum for the $\Lambda(t)$CDM model presents a slightly larger signal as compared to the standard model. This seems to be a virtue of the particle creation model since we have learned from Refs. \citep{giann,crossok,Spergel,stack,stackb,stackc,stackd} that models with higher $C^{Tg}_l$ power are desirable. Nevertheless, given the large uncertainties in determining the observed $C^{Tg}_l$ values, both models remain compatible with data.

\section{Conclusions}

The concordance particle creation model is a viable alternative to the standard $\Lambda$CDM paradigm, with the same free parameters, namely $\Omega_{m0}$ and $H_0$. Since particle creation at a constant rate leads to a different background and perturbative dynamics, it is important to investigate specific signatures of this model in comparison to the standard cosmology.
We provided in this work a direct computation of the evolution of perturbed scalar quantities like the matter and baryonic density contrasts and the gravitational potential for the $\Lambda(t)$CDM model. There is a clear scenario emerging in such a cosmology (see Fig. \ref{deltas}), in which the dark matter contrast is highly suppressed at late times while the baryonic contrast maintains a constant value, as in the standard case, but with a slightly smaller amplitude. This dynamics does indeed lead to a consistent description of the structure formation data \citep{LtCDMbp2008,LtCDMp2011, LtCDMp2015}.
 Our main goal here was to provide a comprehensive analysis of the ISW effect in such concordance particle creation cosmology. With the results for the perturbative dynamics calculated in section II we computed the CMB-LSS cross correlation spectrum. Our care in calculating the evolution of both baryonic and total matter contrasts in detail leads to a full and safe analysis of the $C^{Tg}_l$ spectrum. 

Current efforts \citep{giann,crossok,Spergel,stack,stackb,stackc,stackd} in obtaining the observed $C^{Tg}_l$ spectrum are sending a clear message: the CMB-LSS signal seems to slightly exceed the $\Lambda$CDM prediction.  Therefore, it is timely to check the $\Lambda(t)$CDM outcomes for this cosmological observable.
For both the NVSS (Fig. \ref{fig3}) and WISE (Fig. \ref{fig3b}) data, the $\Lambda(t)$CDM model leads to a desirable excess of power in the $C^{Tg}_l$ spectrum of order $10$-$20\%$. This is in fact a small increase that potentially cannot be sensitive to ISW probes. 
It is worth noting that the available data is insufficient for distinguishing cosmological models with a reliable statistical confidence. This happens mainly because of the cosmic variance limits imposed to large scale CMB analysis, that leads to a very low signal-to-noise ratio. Here we are restricted to a qualitative comparison between the $\Lambda(t)$ and standard cosmologies. Future data can in principle improve the accuracy of the CMB-LSS cross-correlation technique and therefore specific features of different cosmological models concerning the ISW effect could be compared in more detail. Although pure ISW detection techniques do not seem to be a powerful tool in discriminating cosmological models, they can eventually complement other cosmological probes.

\section*{Acknowledgements}
We are thankful to the authors of Ref. \citep{wang} for making their code available and to J. Enander, S. Ilic and C. Pigozzo for helpful discussions. HAB thanks the partial support of CNPq (Brazil), through the ``Science without Border'' program, and Fapesb. SC thanks the partial support of CNPq, grants No. 309792/2014-2 and 479937/2013-3. HV acknowledges CNPq and support of A*MIDEX project (No. ANR-11-IDEX-0001-02) funded by the ``Investissements d'avenir" French Government program, managed by the French National Research Agency (ANR). The work of RF and SG is funded by CNPq.

\label{lastpage}

\end{document}